# Multi-component measurements of the Jefferson Lab energy recovery linac electron beam using optical transition and diffraction radiation[*]

M. A. Holloway, R. B. Fiorito[†], A. G. Shkvarunets, P. G. O'Shea

Institute for Research in Electronics and Applied Physics

University of Maryland, College Park, MD 20742

S. V. Benson, D. Douglas, P. Evtushenko, K. Jordan

Jefferson Lab, Newport News, VA 23606

## Abstract

High brightness electron accelerators, such as energy recovery linacs (ERL), often have complex particle distributions that can create difficulties in beam transport as well as matching to devices such as wigglers used to generate radiation from the beam. Optical transition radiation (OTR), OTR interferometry (OTRI) and optical diffraction-transition radiation interferometry (ODTRI) have proven to be effective tools for diagnosing both the spatial and angular distributions of charged particle beams. OTRI and ODTRI have been used to measure *rms* divergences and optical transverse phase space mapping has been demonstrated using OTRI. In this work we present the results of diagnostic experiments using OTR and ODR conducted at the Jefferson Laboratory's 115 MeV ERL which show the presence of two separate components within the beam's spatial and angular distributions. By assuming a correlation between the spatial and angular features we estimate an *rms* emittance value for each of the two components.

---

[*] work supported by ONR and DOD Joint Technology Office
[†] corresponding author: rfiorito@umd.edu

## I. Introduction

Multiple components are often present in the transverse phase space of charged particle beams. For example, the presence of beam halo and a core have been observed in the spatial distributions of intense electron and proton beams. Understanding the properties of each beam component can be important in the successful operation of the accelerator and for optimizing the performance of beam-radiation devices, such as wigglers for free electron lasers.

Halos are particularly worrisome since they can be a source of unacceptable particle losses which can cause activation or severe damage to the accelerator. Control of halos is crucial to the development of the next generation of high power, high brightness accelerators. While there are studies on halo formation [1- 4], little is known about its origin or how to mitigate its effects. It is therefore important to have diagnostics which are sensitive to such beam components.

Optical transition radiation diagnostics have proven effective for measuring the transverse phase properties of both relativistic and non relativistic electron and protons beams. High resolution OTR imaging is commonly used to profile the beam and OTR interferometry (OTRI) and optical diffraction-transition radiation interferometry (ODTRI) have both been successfully used to measure the divergences of relativistic beams [5,6]. In addition, OTRI has been used to optically map the transverse phase space of a relativistic beam [7].

Previous experiments using OTRI and ODTRI have both shown evidence of distinct components within a beam's *angular* distribution [5]. However, these measurements were not sensitive enough to show the presence of distinct *spatial distributions* within the same beam.

In this study we present the results of OTRI and ODTRI measurements of the Jefferson Lab ERL, which simultaneously indicate the presence of two spatial and angular distributions, i.e. a core and halo component, within the beam. By assuming a correspondence between the core spatial and the low divergence angular component, and the halo with the higher divergence component, we have estimated the rms emittances of the two components and compared the core emittance to previous independent measurements at JLAB.

## II. Theoretical Background

### A. Optical Transition Radiation Interferometry

A conventional OTR interferometer consists of two parallel metal foils oriented at 45 degrees with respect to the beam direction; the first foil is usually a solid thin metal and the second foil is a mirror. Forward OTR is produced as the electrons emerge from the first foil and backward OTR is produced as the electrons enter the second foil. The interfering forward and backward OTR interfere and are observed in reflection at the angle of specular reflection, i.e. at 90 degrees with respect to the direction of the electron beam. The position and visibility of these fringes can be used to measure beam energy, energy spread and divergence [8].

To obtain sufficiently numerous fringes for a divergence measurement, the distance is between the foils must be greater that the vacuum coherence length, $L_v = \lambda/[\pi(\gamma^{-2}+\theta^2)]$, which is defined as the distance required for the phase of the field of the electron and the OTR photon that it generates to change by $\pi$ radians. Here $\lambda$ is the observation wavelength, $\gamma$ is the Lorentz factor, and $\theta$ is the angle of observation which, for forward OTR, is measured from the direction of the velocity vector of the beam and, for backward OTR, from the direction of specular reflection.

The spectral-angular distribution of intensity produced by a single relativistic charge passing through two parallel metal foils of an OTR interferometer is given by

$$\frac{d^2 I}{d\omega d\Omega} = \frac{q^2}{\pi^2 c} \frac{\theta^2}{(\gamma^{-2}+\theta^2)^2} \sin^2\left(\frac{\varphi}{2}\right) \qquad (1)$$

where $q$ is the charge and $\varphi = L/L_V$ is the difference in phase between the OTR generated at the first foil and the OTR generated at the second foil and $L$ is the foil separation. The first term in Eq. (1), the single foil OTR angular pattern is a slowly varying function of angle and energy but the phase and hence the interference term of Eq.

(1) is a rapidly changing function of these variables as well as the wavelength of observation. Thus the visibility of the interference fringes is highly sensitive to the electron beam's energy, energy spread and divergence as well as the inter-foil spacing and observational band width.

**B. Optical Diffraction-Transition Radiation Interferometry**

For beams with low energy and or low divergence, a conventional OTR interferometer is limited in its ability to measure divergence by beam particle scattering in the front foil. To overcome this limitation we have used a micromesh foil to replace the solid foil. When electrons or other charged particle pass through such a mesh optical diffraction radiation is generated from particles passing through the holes as well as the wires of the mesh when the condition $\gamma\lambda/2\pi \geq a$, where $a$ is the size of the hole or wire. With the right combination of mesh and experimentally controllable parameter, e.g. wavelength and band pass, the ODR generated from the unperturbed particles passing through the holes will form interferences when coherently added to OTR generated by the same particles intercepting the mirror, and the interferences from particles scattered in the mesh wires will be washed out forming a smooth background above which the interferences from unscattered particles will be visible. The details of the design and operation of an ODTR interferometer have been presented in Ref. [5]. Because of the complexity of the interactions, a simple formula such as Eq. (1) is unavailable to present ODR from the mesh, corresponding to the slowly varying envelop function in Eq. (1) and we have developed a simulation code developed to predict it [5]. However, since the difference in phase between ODR or OTR photons generated from either mesh or a foil and OTR from the mirror is identical, the fringe function, i.e. the sine term of Eq. (1), is identical and thus the analysis of the fringe visibility presented below applies equally well to OTRI and ODTRI.

**C. Effects of Divergence, Energy spread and Optical Band Pass on OTRI/ODTRI**

The effects of divergence, energy spread, and bandwidth on the fringe visibility can be estimated by taking the total variation of the phase difference between the OTR or ODR generated in the first foil of either interferometer and the OTR from the mirror.

The details of this analysis have been presented previously by us [8] so we only present a summary of the results here.

The relative phase between two photons generated by a single electron with trajectory angle $\theta_e$ from two foils in its path which are separated by distance $L$, and observed at an angle $\theta$, for $\theta, \theta_e \ll 1$ and $\gamma \gg 1$, is given by [8]:

$$\psi = \frac{\pi L}{\lambda}\left(\gamma^{-2} + \theta^2 - 2\theta \cdot \theta_e\right) \tag{2}$$

Note that this phase difference is different from the phase $\varphi$ defined following Eq. (1) i.e. $\Psi$ contains the additional term $(-2\theta \cdot \theta_e)$, to account for the possibility of a non zero particle trajectory angle.

An estimate of the spread in relative phase $\Delta\psi$ caused by the observational bandwidth ($\Delta\lambda$), the beam energy spread ($\Delta\gamma$), and the beam divergence ($\Delta\theta_e$) can be calculated by taking total variation of the relative phase. When this is done and we choose $\Delta\theta = 0$, i.e. a fixed angle of observation, and $\theta_e = 0$, i.e. propagation of the beam along the axis,

$$\Delta\psi = \frac{2\pi L}{\lambda}\left[\theta^2 \frac{\Delta\lambda}{\lambda} + \theta \cdot \Delta\theta_e + \gamma^{-2}\frac{\Delta\gamma}{\gamma}\right]. \tag{3}$$

From Eq. (3) we can estimate how much a given energy spread $\Delta\gamma$, angular divergence, $\Delta\theta_e$ and filter bandwidth, $\Delta\lambda$ will affect the visibility of the interference fringes. It is clear that if $\Delta\Psi = 0$ the fringes are 100% visible and that an increase in $\Delta\Psi \geq 2\pi$ results in a complete washing out of the fringes. We can then heuristically infer that when $\Delta\Psi \simeq \pi$, the fringe visibility will be about 0.5 and most sensitive to a change or spread in parameters.

However, a more quantitative estimate of the effect of a spread in any of the parameters in Eq. (3) on the fringe visibility can be calculated as follows. The fringe intensity modulation, which is explicitly given in Eq. (1), can be presented in the form $\tilde{I} = (1 - A\cos\Psi)/2$, where the coefficient $A$ is the fringe visibility. For a no spread in parameters $A = 1$ and in the presence of a spread we can expect that $0 < A < 1$. In order to calculate *A,* we assume that the phase interval $\Delta\Psi$ is associated with an ensemble of radiators with phases distributed within this interval. Then the elemental intensity of radiation associated with phase $\Psi+\xi$, within the interval $\Delta\Psi$ is given by $d\tilde{I}=D(\xi)\cdot[1-\cos(\Psi+\xi)]d\xi/2$, where $D$ is a weighting function which depends on $\xi$, the phase shift from the point $\Psi$, which we assume is the center of the interval $\Delta\Psi$. We also assume for simplicity that the weighting function can be represented by a Gaussian distribution, i.e. $D(\xi,\sigma)=\pi^{-1/2}\sigma^{-1}\exp(-\xi^2/\sigma^2)$, where $2\sigma$ is the full width of the distribution and $\Delta\Psi=2\sigma$. Then the total fringe intensity can be written:

$$\tilde{I} = \frac{1}{2\sigma\sqrt{\pi}} \int_{-\infty}^{\infty} [1-\cos(\Psi+\xi)]\exp(-\frac{\xi^2}{\sigma^2})d\xi = \frac{1}{2}[1-\exp(-\frac{\sigma^2}{4})\cos(\Psi)] \qquad (4)$$

Note that the term $\exp(-\sigma^2/4)$ is the just the fringe visibility *A*, which is now determined in terms of the width of the distribution and, equivalently, the phase spread $\Delta\Psi$. The maximum change of the fringe visibility with respect to a change in width can be easily calculated from Eq. (4) and occurs at $2\sigma=2.83$, a value close to our original guess, i.e. $\Delta\Psi\simeq\pi$, where the visibility *A* = 0.6. The fringe visibility drops down to 0.1 at $2\sigma=4.3$ and to 0.01 at $2\sigma=6$.

We can also directly conclude from Eq. (3) that in all cases the fringe visibility decreases as *L/λ*, increases. Furthermore, if we set $\Delta\theta_e=\Delta\lambda=0$, we see that the effect of energy spread, Δγ on the fringe visibility is independent of observation angle. Similarly if Δγ and Δλ are neglected, the effect of divergence is proportional to the observation angle but is independent of the energy, γ. Finally, the effect of bandwidth is proportional to the square of observation angle but is independent of the beam energy. These dependences can be used to advantage to diagnose either the energy spread or the divergence but in any case control of the bandwidth is necessary. Experimentally we can adjust *L* and λ to optimize the number of fringes for a given range of divergence or energy spread and chose a narrow band pass filter to minimize the effect of the band pass on the visibility.

If the beam energy spread is small compared to the divergence and a sufficiently narrow band pass filter is used, the effect of divergence will dominate the fringe visibility. Under these conditions the visibility of OTR or ODTR interferences is a divergence diagnostic [6] and the position of each the fringes is a diagnostic for the average energy of the beam particles.

Previous measurements of the JLAB ERL beam have shown that the fractional energy spread $\Delta\gamma/\gamma \approx 0.02$ and that the expected normalized rms beam divergence $\Delta s = \gamma\Delta\theta_e$ is in the range 0.05 - 0.1. Table 1. shows the effects of each of the terms of Eq. (3), calculated using the above estimates for a 2% energy spread and a 2% bandwidth (10nm at 650nm and 450nm) for the lowest and highest divergences expected. The results show that the effect of energy spread is negligible for the entire range of observation angles but that there are possible competing effects from bandwidth at the larger observation angles for the lower value of divergence (0.221 mrad, $\Delta s$ = 0.05). The normalized observation angle $\gamma\theta$, i.e. the observation angle measured in units of $1/\gamma$, is presented in the first column of Table 1.

| Normalized angle of observation | Angle of observation (mrad) | Effect of 2% band width filter | Effect of 0.221 mrad divergence | Effect of 0.442 mrad divergence | Effect of 2% energy spread |
|---|---|---|---|---|---|
| 1 | 4.42 | 0.136 | 0.442 | 0.882 | 0.18 |
| 2 | 8.84 | 0.545 | 0.882 | 1.764 | 0.18 |
| 3 | 13.27 | 1.224 | 1.323 | 2.646 | 0.18 |
| 4 | 17.68 | 2.176 | 1.764 | 3.528 | 0.18 |

**Table 1: Variation of phase terms calculated for the JLAB experimental parameters.**

We note, however, that Eq. (3) can only provide an approximation of the relative effects of variations in the parameters since it does not account for distributions of the parameters. For a more exact analysis, convolutions of the intensity, e.g. Eq. (1) for OTRI, with each of these distributions is needed. When this is done, using even simple 1D models for the distribution functions, e.g. a single Gaussian distribution of particle angles and a rectangular filter function, it is seen that the effect of a 2% bandwidth filter is small for the entire range of observation angles but that a larger bandwidth obscures the divergence effect at the larger angles of observation.

## D. Two Dimensional Convolution of OTRI and ODTRI

In order to provide an firmer basis to fit the real interferometric data, which is two dimensional, we have performed 2D angular convolutions of the calculated OTRI far field angular pattern, i.e. Eq. (1), and the computer simulated ODTRI far field pattern using a computer code [5,7], with a sum of up to three, two dimensional Gaussian components to model the distribution of the particle trajectory angles. The convolution of such a distribution with the spectral angular distribution of intensity for OTRI or ODTRI produces an interference pattern whose fringe visibility is sensitive to the *rms* widths of the individual Gaussian components. In addition, convolutions of a rectangular filter function and a cosine distribution of energy are also performed.

The convolved patterns are then analyzed to produce vertical and horizontal line scans that are fit to those obtained from the measured OTRI/ODTRI intensity patterns. To perform the fit, horizontal and vertical sector scans ($\theta_x$ and $\theta_y$) are first obtained for each OTRI/ODTRI interference pattern. The convolution code calculates horizontal (*x*) and vertical (*y*) intensity profiles of the single electron intensity distribution, e.g. Eq. (1) for OTRI, with the angular distribution model. The variances of the model distribution are then adjusted manually to simultaneously fit the horizontal and vertical sector averaged line scans obtained from the data. The best fit is achieved by minimizing $D(\alpha)$ the integral *rms* deviation between the calculated curve and the data which is defined [6] as:

$$D(\alpha) = \frac{1}{(\theta_2 - \theta_1)} \left\{ \int_{\theta_1}^{\theta_2} \left( \frac{\alpha \cdot E(\theta) - T(\theta)}{\alpha \cdot E(\theta) + T(\theta)} \right)^2 d\theta \right\}^{\frac{1}{2}} \tag{5}$$

where *T(θ)* is the value of the calculated curve at observation angle *θ*, *E(θ)* is the value of the experimental data at *θ* and *α* is an arbitrary scaling constant. In the fitting procedure, the variable *α* is varied until *D(α)* is minimized.

## E. RMS Emittance

The normalized rms emittance for a given dimension $r = x, y$ is given by [10]:

$$\tilde{\varepsilon}_r = \gamma\beta(\langle r^2\rangle\langle r'^2\rangle - \langle rr'\rangle^2)^{\frac{1}{2}} \qquad . \qquad (6)$$

At the waist of the beam envelope the correlation term <rr'> is zero and the normalized emittance reduces to

$$\tilde{\varepsilon}_r = \gamma\beta r_{rms} r'_{rms} \qquad (7)$$

Where $\beta = v/c$, $r_{rms} = \sqrt{\langle r^2\rangle}$, and $r'_{rms} = \sqrt{\langle r'^2\rangle}$. We take simultaneous measurements of both of these quantities at both horizontal and vertical waists and use Eq. (7) to compute the corresponding horizontal and vertical rms emittances.

### III. Experimental Setup

#### A. Interferometer Design

Both OTR and ODTR interferometers are used in this study. Their designs are similar to those used in previous experiments [6,8]. Both interferometers are mounted on a ladder which can be sequentially lowered into the beam path by means of a linear actuator. The front foil of the OTRI interferometer is a solid, thin aluminum foil which is 0.7 μm thick. Our calculations show that at a beam energy of 115 MeV the normalized rms scattering angle in this foil $\gamma\theta^{scat}_{rms} < 0.02 < \Delta s$ so that this foil is not expected to influence the measured divergence. The first foil of the ODTRI interferometer is a nickel micromesh which is 5μm thick and has square holes 11.2 μm in width, spaced with a period of 16.9 μm. The common second foil of both interferometers is an optically flat, rectangular piece of silicon cut from a standard wafer which is 0.5 mm thick. The silicon is coated with 1000 Å of aluminum to form an optical mirror with better than 90% reflectivity in the visible. The mirror surface is parallel to the front surface of the OTR foil and ODR mesh and the spacing between the front faces of the foil and mesh and the mirror surface L = 47 mm.

The ladder also contains section a below the interferometers exposing the mirror, which is used for optical alignment as well as beam size measurements, and a standard 19 mm diameter circular optical graticule, whose surface is coplanar with the mirror surface.

Aluminum crosshairs vacuum deposited on the surface of the graticule form a target used to determine the magnification of the beam imaging optics. The crosshairs have 10 major divisions on each axis; each division is 0.5mm in length.

The ladder assembly is housed in a standard 152 mm, six way vacuum cross with fused silica windows which is located just before a beam dump as shown in the upper left part of Figure 1. During experimental runs the electron beam is switched into the interferometer and beam dump line instead of the magnetic bend, which is the normal operation of the ERL system.

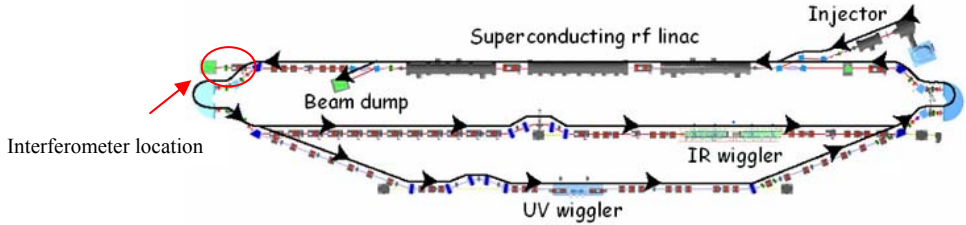

**Figure 1: Jefferson Lab FEL ERL**

**B. Optics**

The optics used in our experiments were designed to carry out future phase space mapping experiments that require a magnified image of the beam at a secondary focal plane. However, they served well to perform the *rms* measurements described in this paper. The optics are arranged on 70 by 140 cm optical breadboard, which is at the same level as the beam line. Figure 2. is an overhead schematic of the optics table. The input light emerges from a side port of the vacuum cross and onto the optical breadboard on the upper right corner of Figure 2. The red and green lines represent the optical paths of the rays used for near field (beam) and far field imaging, respectively. As is clear from the diagram both paths coincide until they reach the pellicle beam splitter, which reflects 10% of the light into the path of the beam or near field imaging system, while 90% passes through the splitter to the angular distribution or far field imaging system.

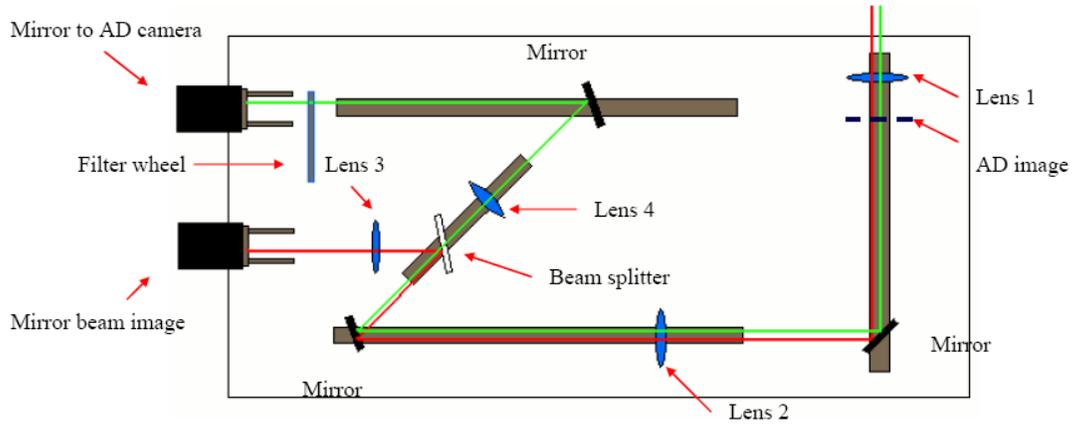

**Figure 2: Optical Layout**

Lens 1 and 2 have focal lengths 200 mm and 100 mm, respectively, and are spaced 720 mm apart. Together they create a ten times magnified image of the beam on the surface of the beam splitter, an uncoated pellicle with 90% transmission in the visible. Lens 3 has a focal length of 400 mm and transports the beam image to the beam imaging camera. An image of the far field angular distribution of OTRI or ODTRI is created at the focal plane of Lens 1. Lens 4, with a focal length of 400 mm, relays the AD image to the AD camera. The filter wheel introduces a 650 nm x 10 nm band pass filter, a 450 nm x 10 nm filter or a clear aperture into the angular distribution light path.

The beam dump is very close to the experimental setup and is a source of high energy X-ray radiation. Thus the CCD cameras must be shielded with lead to reduce the image noise produced by x-radiation and to protect them from damage. To make shielding easier, both cameras are placed near the floor which is about 92 cm below the level of the beam dump. The light from both image paths is directed toward the floor by the two mirrors shown on the far left of Figure 2. Figure 3. is a side view of the light paths leading to both cameras. Lenses 5, and 6, which have focal lengths 200 mm and 100 mm respectively, focus the images formed on the breadboard onto the shielded CCD cameras shown in Figure 3.

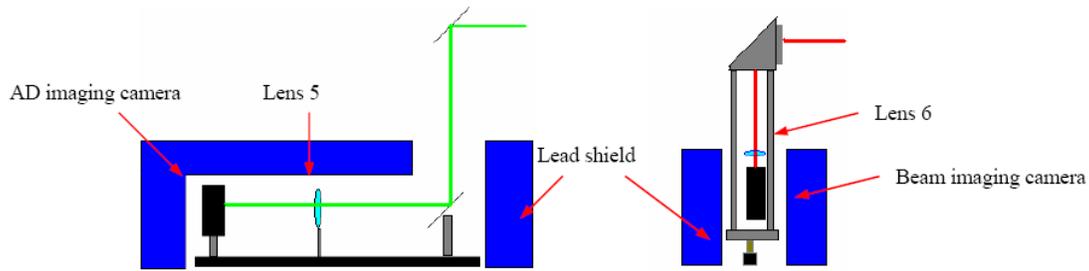

**Figure 3 : Side view of light paths to cameras; top rays are outputs from the breadboard optics shown in Figure 2.**

Lenses 1 – 5 are all doublet achromatic lenses and Lens 6 is a standard Nikon (135mm, f/2.8) camera lens. Achromats throughout the optical train are necessary to minimize both spherical and chromatic aberration. The entire optical system is designed to ensure an acceptance angle of $10/\gamma$. From simulations, the interference fringes are expected to be observable out to an angle of about $5/\gamma$. We maintain an acceptance angle of $10/\gamma$ throughout the optical system to insure that no light rays of interest are lost in the optical transport.

Ray transfer matrices were used to calculate the size of the light ray bundle throughout the entire optical path in order to optimize the throughput and to insure that the required angular acceptance is maintained throughout the optical paths. Using thin lens approximations, ray transfer matrices can be used calculate the height from the optical axis and angle with respect to the optical axis of a light ray at any point in an optical system with a given input height and angle [10]. The largest electron beam radius expected at the interferometer is about 1 mm. The height and angle of the ray is then checked at the surface of every lens in the entire system to ensure the lens will capture the ray. The same calculation is performed for the next lens surface in the optics train using the new height and angle at the first lens. This process is repeated until the surface of the camera sensor is reached. Figure 4. shows the results of ray transfer matrix calculations performed in MATLAB for both the far field and near field beam paths.

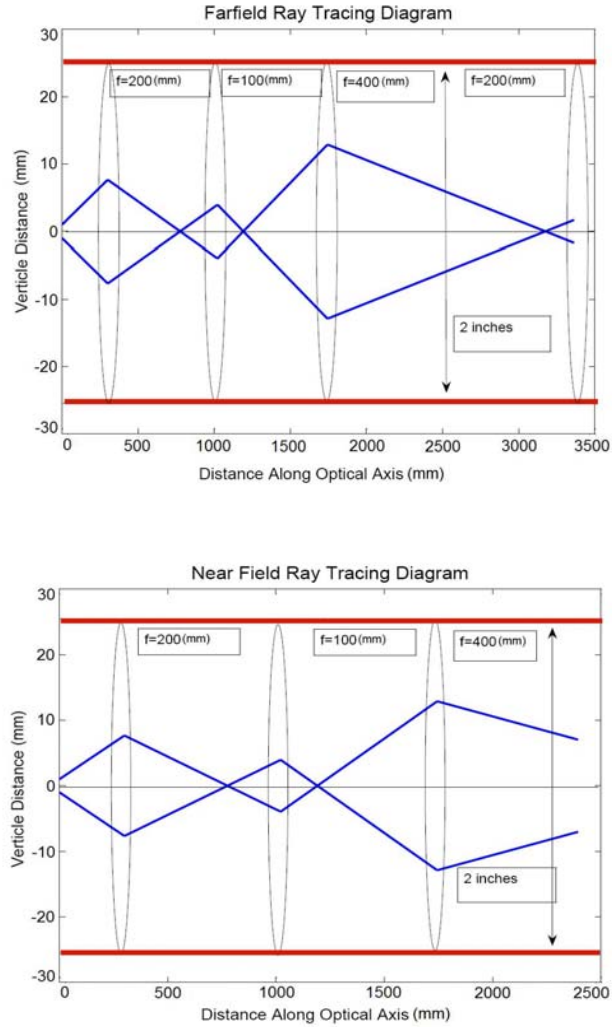

**Figure 4: Ray tracing plots for the near field and far field optical path**

The initial rays shown in blue start at height of 1 mm from the optical axis and have angles $\pm 10/\gamma$ with respect to the optical axis. The minimum aperture in the optics system is 50mm in diameter which is represented by the red lines in Figure 4.

**C. Optical Alignment**

To align the optics a HeNe laser installed near the electron gun of the ERL is used. The laser beam travels down the full length of the LINAC from the injector through the beam pipe along the electron beam path. The laser spot is about 20 mm diameter when it reaches the interferometer. The ladder is then adjusted so that the nickel mesh position is in the laser's path.  When the laser strikes the nickel mesh a diffraction pattern is created which is reflected from the silicon mirror into the optical system. Figure 5. is the image of the Fraunhofer diffraction pattern observed by the far field camera.

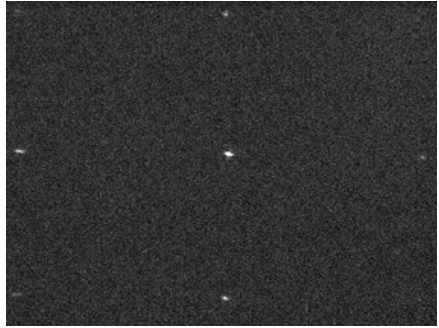

**Figure 5: Diffraction pattern from the mesh as observed by the far field camera**

The diffraction pattern consists of a rectangular pattern of spots; the central spot contains light at an angle of zero degrees with respect to the electron beam axis. The field of view of the camera is such that only the central and first order dots are seen in Figure 5.  The central order of the diffraction pattern serves as a reference spot to align the optics. In this procedure, the lenses are initially removed. The mirrors are adjusted so the central order laser beamlet travels along both the near field and far field beam paths at a constant height.  Each lens is then placed in its proper location and adjusted so that the laser spot travels through the center of each lens.

**D. Far Field Camera**

The far field camera is a high quantum efficiency, 16 bit, cooled CCD camera (SBIG model ST-402ME) which is commonly used for astronomical observations.  The CCD sensor array consists of 765 x 510, 9 micron square pixels. The camera is computer controlled and acquires a single image over a specified integration time. The exposure time is controlled by a mechanical shutter and allows integration times from 0.04 to 3600 seconds. The images are downloaded via a USB 2 link.  The SBIG camera must be heavily shielded from radiation in all direction due to its sensitivity.  Lead bricks completely enclose the camera with at least 100 mm of lead.

The far field camera is also focused using the diffraction pattern described above by adjusting Lens 4 and 5 until the diffraction spots are sharpest. The calibration and angular field of view are determined by calculating the angular position of the first order diffraction spots located directly horizontal and vertical from the central order. Each spot is at an angular position $\theta = \lambda/d$ where $\lambda$ = 632 nm, the HeNe alignment laser wavelength, and $d$ = 16.9 μm is the period of the micromesh. Dividing the angular spacing of the first order spots by the number of pixels between the central order and the first order provides the angular calibration for the far field camera. The 45 degree tilt of the interferometer foils with respect to the beam axis decreases period seen by the laser in the horizontal direction shorter by the factor $\sqrt{2}$ and is the reason why the horizontal first order spots are observed to be a greater distance from the central order than the vertical spots.

Calculating the field of view is important to ensure there are enough pixels to resolve each interference fringe. The angular field of view in the vertical direction of Figure 4. is about $15/\gamma$. Simulations show that about 6 fringes are expected out to a an angle of $3/\gamma$. The total number of pixels in a vertical line in our CCD camera is 510. With an estimated 6 fringes covering 1/5 of the pixels in the vertical direction this provides about 17 pixels per fringe, which is quite acceptable for our fringes visibility measurements.

**E. Near Field Camera**

The near field camera is a standard RS-170 video CCD camera which is the standard used by the Jefferson Lab FEL to monitor the OTR generated by electron beam at various pop in foils along the beam line. The camera feed is attached to a 10 bit frame grabber and image acquisition is synchronized to the drive laser pulse of the electron gun.

To focus the near field (beam imaging) camera to the surface of the mirror surface of the interferometer, the ladder is moved to the graticule position. The alignment HeNe laser is then used to illuminate the graticule at the bottom of the ladder and thus to create an image on which to focus the beam imaging camera. Lenses 1 and 2 are adjusted to focus the image of the crosshairs on the graticule, which is visible to the naked eye on the surface of the beam splitter and then Lens 6 is adjusted to focus the near field camera onto the image at the surface of the beam splitter. Calibration is achieved by measuring the number of pixels per division of the crosshairs.

**F. Operating Conditions**

In order to focus the beam to either a horizontal or vertical waist condition at the mirror of the interferometer we use 3 quadrupole pairs which are located upstream of the interferometer. Simultaneous images are then obtained of the far field OTR or ODTR interference pattern along with the beam's spatial intensity distribution produced at the surface of the second foil (mirror) of the interferometer.

Each far field OTR or ODTR interferograms is obtained in a 90 second exposure with one of two filters in place, a 650x10nm or a 450x10nm band pass filter. An integration time longer than 90 seconds lead to increased image noise due to X-rays and did not significantly improve the signal to background ratio. While background pictures were obtained with the camera shutter closed for the same time duration with the beam on, it was found that subtraction of the background image did not significantly improve the signal to noise and in some cases produced negative intensity values. Therefore the raw images were used for data analysis.

The accelerator beam conditions for our experiments are listed in Table 2.

| Beam energy | 115 MeV |
|---|---|
| Macro pulse width | 100μs |
| Micro pulse rep rate | 2MHz |
| Charge per bunch | 135 pC |
| Beam Current (avg) | ~150μA |

Table 2: Experimental beam conditions

## IV. Results

### A. Divergence Measurements

Figure 6. shows far field OTRI interference pattern taken at a vertical (left) and horizontal (right) waist conditions with a 650 x10nm band pass filter. The color overlays are the sectors used to average the pixels at each radius to produce vertical and horizontal line scans, respectively.

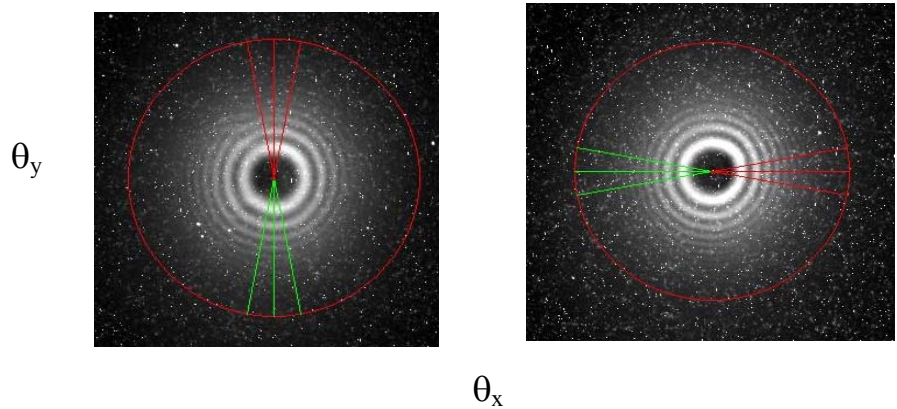

$\theta_y$

$\theta_x$

**Figure 6: OTRI patterns taken with a 650 x 10nm band pass filter at a vertical waist (left) and a horizontal waist (right) with overlays of sectors used to produce averaged vertical and horizontal line scans.**

A good theoretical fit of the intensity of vertical or horizontal scans of OTRI and ODTRI could not be achieved assuming a single Gaussian component for the vertical and horizontal angular distributions of the particles at either waist condition. To get a good fit at least two Gaussian components were needed. Figures 7. and 8. show the fits to the vertical and horizontal scans of the OTRI and ODTRI patterns obtained with two optical band pass filters, at the Y and X waist conditions, respectively. The legend in each plot shows the half widths ($\sigma_1, \sigma_2$) of the two Gaussian functions used in the fit.

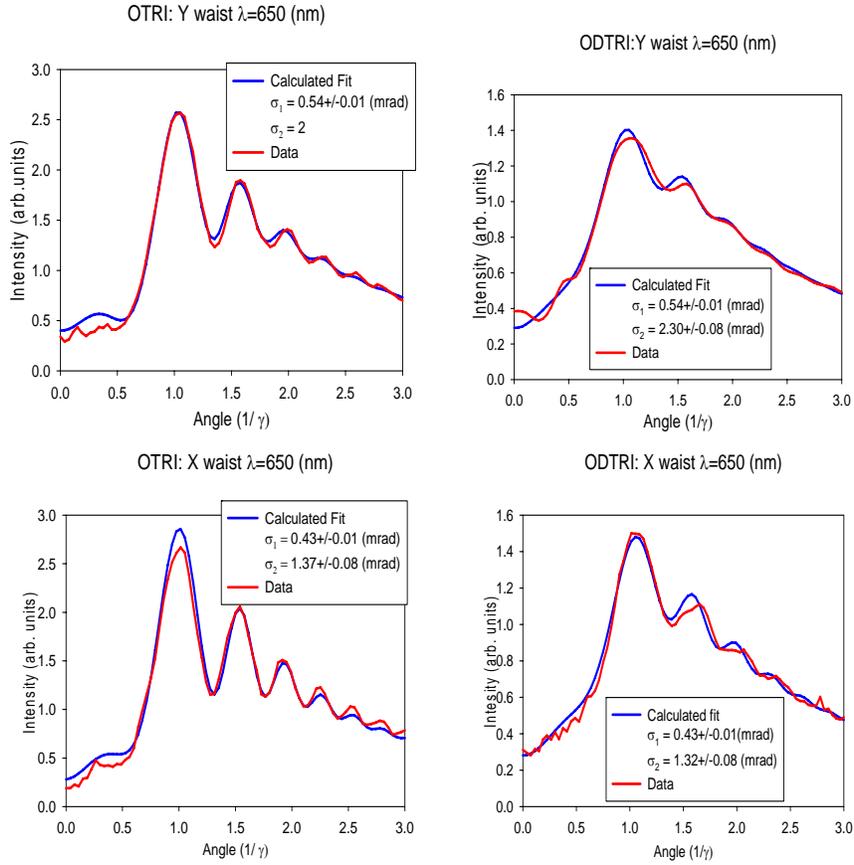

**Figure 7: OTRI and ODTRI fits and data taken with a 650 x 10nm optical band pass filter at vertical (Y) and horizontal (X) beam waists.**

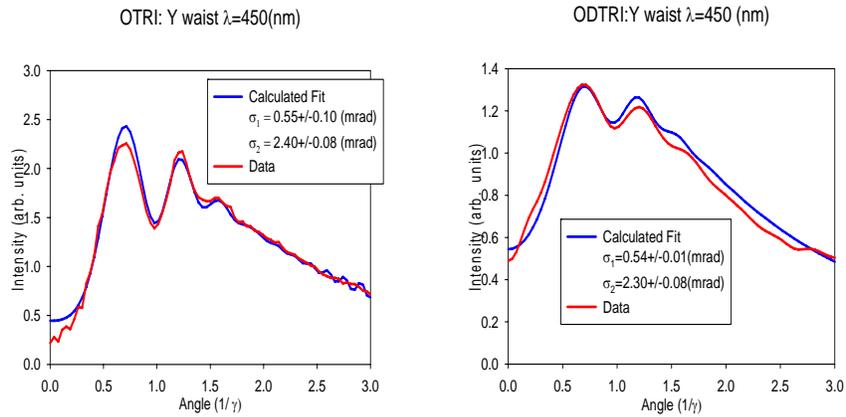

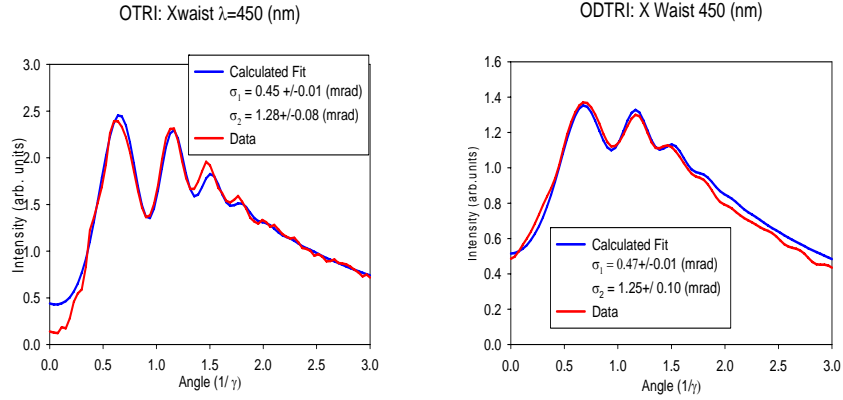

**Figure 8: OTRI and ODTRI fits and data taken with a 450 x10 nm band pass filter at vertical (Y) and horizontal (X) beam waists.**

The rms divergences and the rms deviation $D(\alpha)$ for the two Gaussian fits to the OTRI and ODTRI data for each waist are listed in Table 3. The error in each divergence component listed in the table is an indication of the sensitivity of the overall fit; i.e. a change in a given divergence component by the error listed, holding the other component constant, produces a change in $D(\alpha)$ of 0.01%.

We note that the vertical and horizontal divergences are comparable but that the vertical divergence is about 20% larger than the horizontal. The excellent agreement between the divergences obtained with OTRI and those obtained with ODTRI confirms our initial prediction that scattering in the OTR interferometer's first foil has no significant effect on the divergence measurements using OTRI.

| Waist | Method | Filter (nm) | Vertical Divergence 1 (mrad) | Vertical Divergence 2 (mrad) | $D(\alpha)$ |
|---|---|---|---|---|---|
| Y | OTRI | 650x10 | 0.54+/-0.01 | 2.30+/-0.08 | 3.23% |
| Y | ODTRI | 650x10 | 0.54+/-0.01 | 2.30+/-0.08 | 1.33% |
| Y | OTRI | 450x10 | 0.55+/-0.01 | 2.40+/-0.08 | 4.25% |
| Y | ODTRI | 450x10 | 0.54+/-0.01 | 2.30+/-0.08 | 2.43% |
| X | OTRI | 650x10 | 0.49+/-0.01 | 1.59+/-0.08 | 5.18% |
| X | OTRI | 450x10 | 0.45+/-0.01 | 1.56+/-0.08 | 3.75% |
|   |   |   | Horizontal Divergence 1 (mrad) | Horizontal Divergence 2 (mrad) |   |
| X | OTRI | 650x10 | 0.43+/-0.01 | 1.37+/-0.08 | 5.42 % |
| X | ODTRI | 650x10 | 0.43+/-0.01 | 1.32+/-0.08 | 2.73% |
| X | OTRI | 450x10 | 0.45+/-0.01 | 1.28+/-0.10 | 5.39% |
| X | ODTRI | 450x10 | 0.47+/-0.01 | 1.25+/-0.10 | 1.79% |

**Table 3:** *RMS* divergences from double Gaussian fits to OTRI and ODTRI patterns obtained at vertical (Y) and horizontal (X) beam waists.

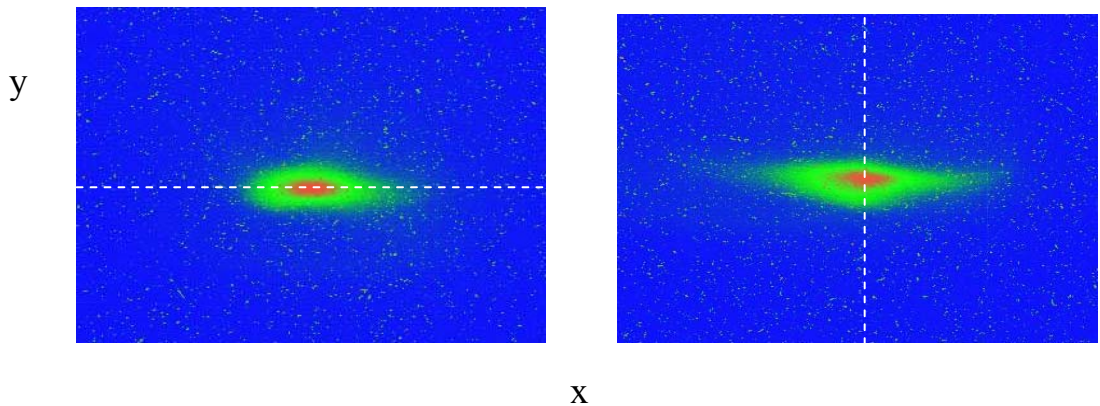

**Figure 9:** Beam images at a horizontal minimum (left) and a vertical minimum (right) showing the horizontal and vertical lines where scans were performed.

## B. Beam Size Measurements

Typically the beam images also show two distinct components. Figure 9. shows OTR images of a beam focused to a horizontal waist (left) and a vertical waist (right). Both images show an intense core distribution surrounded by a lower intensity halo distribution.

We obtained intensity profiles by taking a vertical or horizontal line scans across the x or y centroid of each beam image (see dotted lines in Figure 9) and recorded the pixel values at each position along the line. Up to ten images of the beam were taken for each waist condition. Line scans are taken from each picture at the same location, i.e. across the center of the pattern, and averaged to reduce the error caused by background radiation noise. The uncertainty in the intensity at each pixel value is estimated by calculating the standard deviation from the mean for each intensity value. Figure 10. shows averaged intensity profiles for X (left) and Y (right) waist conditions respectively.

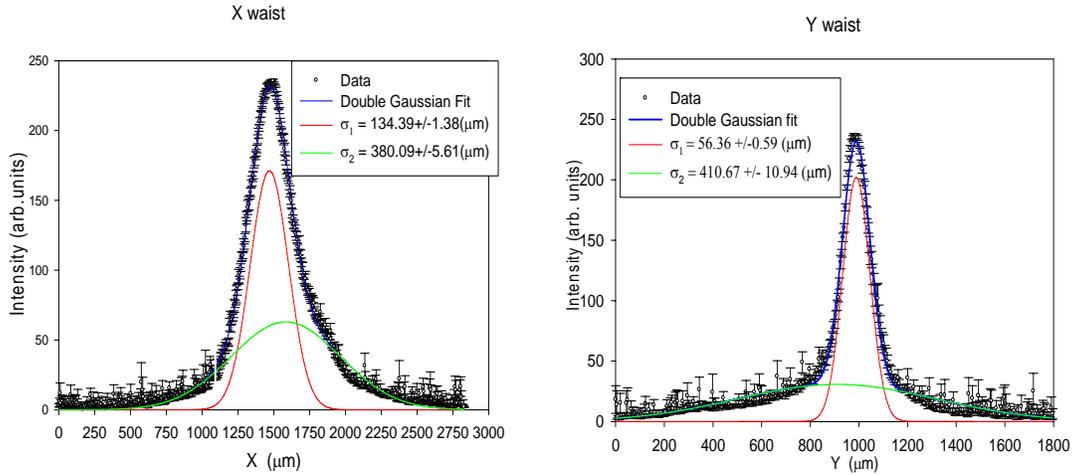

**Figure 10: Averaged line scans at a horizontal waist (left) and a vertical waist (right) along with double Gaussian fits.**

We use the width, σ of each Gaussian component to estimate the vertical and horizontal beam size of the core and halo distributions. The results of the measurements are presented in Table 4. What is noteworthy about the results is: 1) the rms horizontal (x) size of the core for the horizontally focused beam is 2-3 times higher than the rms vertical (y) size of the core for the vertically focused condition; 2) the *y* size of the horizontal focused beam is actually smaller than the *y* size of the vertically focused beam; and 3) the *x, y* sizes of the halo components are comparable for both focusing conditions.

| Waist | Core size (μm) | Halo size (μm) |
|---|---|---|
| X (horizontal scan) | 134.39+/-1.38 | 380.09+/-5.61 |
| Y (vertical scan) | 56.36+/-0.59 | 410.67+/-10.95 |
| X (vertical scan) | 46.17+/-0.61 | 375.04+/-9.42 |

**Table 4. RMS core size and halo size from double Gaussian fits to line scans of beam images taken at X and Y beam waists.**

## C. Emittances calculations

From the experimental data alone we have no way to unambiguously associate either of the two divergences with the core or halo components observed in the spatial distribution of the beam. However, if we make the assumption that the lower divergence component is associated with the core beam distribution and the higher divergence component with the halo, we can estimate the core and halo rms emittances. These are presented in Table 5. The calculation of emittance is done using divergences obtained from OTRI data alone since Table 3. shows that the ODTRI divergence results are very close to those obtained using OTRI. In the first column of Table 5. the symbol X or Y refers to the horizontal or vertical waist and the brackets, i.e. horizontal or vertical scan, refers to the direction of the line scans done to measure the beam size and divergence for each waist condition. Note that the first row of the table provides the horizontal emittance calculated from horizontal line scans of both the near field OTR images and the far field OTRI data; the rest of the rows show vertical emittances calculated from the vertical scans of the OTRI and vertical line scans of the near field images.

In Table 5. there is no entry for the X waist, horizontal emittance for 450 nm, because an experimental error was made in the measurement of the beam size at the time the OTRI far field measurement at this wavelength; this error rendered the emittance calculation for 450 nm invalid. Note that the larger of the two OTRI vertical divergence values measured at an X waist (i.e. the 650nm entry) from Table 3. and the vertical (y) sizes of the core and halo beam components from Table 4. were used to calculate the corresponding vertical emittances, i.e. the last two rows of Table 5., to provide upper bounds on the vertical emittances calculated using vertical divergence measured at an X waist.

| Waist | Emittance | Filter (nm) | Core emittance (mm-mrad) | Halo emittance (mm-mrad) |
|---|---|---|---|---|
| X | horizontal | 650 x 10 | 13.00 +/-0.43 | 117.20 +/- 7.72 |
| Y | vertical | 650 x 10 | 6.80 +/-0.20 | 212.50 +/- 14.89 |
| Y | vertical | 450 x 10 | 6.00 +/-0.23 | 205.40+/- 14.85 |
| X | vertical | 650 x 10 | 5.10 +/-0.17 | 134.20 +/- 10.11 |
| X | vertical | 450 x 10 | 4.60 +/-0.21 | 124.20 +/-10.57 |

Table 5. RMS horizontal and vertical emittances for the core and halo components measured at X and Y waist conditions.

**V. Discussion**

The vertical emittance that is obtained by assuming the core spatial distribution is matched with the lower divergence component agrees with previously estimated values by JLAB of 5-7 mm-mrad. JLAB also estimated that the horizontal emittance was comparable to the vertical emittance. These estimates were obtained with multiple screen measurements and matching the core beam to the FEL wiggler ignoring the beam halo. Our results indicate that the $x$ emittance is about twice the $y$ emittance. Since the measured $x$ and $y$ divergences are comparable, the difference is primarily due to the measured $x$ size of the beam, which is about twice as large as the $y$ size of the beam. The fact that the $y$ size measured at the horizontal waist is actually 20% smaller than the $y$ size measured at the vertical waist indicates that the vertical waist was not actually achieved in the attempt to focus to a vertical minimum.

In both experiments, the vertical and horizontal waists conditions were prescribed visually by the accelerator operator using the OTR image of the beam at the mirror to minimize the *x* or *y* size of the beam at the site of the interferometer mirror in either the *x* or *y* direction. To do this the size of the beam image on the near field camera was monitored. No other means to confirm the actual attainment of a waist was available. Therefore, we can only claim that the measured beam size values and corresponding emittances are upper bounds.

Furthermore, it is quite possible and even likely that the waist condition for the core is not the same as that for the halo. In these experiments no attempt was made to minimize the size of the halo in either the *x* or *y* direction. This was impractical to do due to the limited dynamic range of our beam imaging camera and the intensity of the core component. Thus, we can only claim upper bounds for the halo emittances as well.

## VI. Conclusions

We have shown that an analysis of spatial and angular distributions of OTR (or ODR), simultaneously observed from the 115 MeV JLAB ERL electron beam, each show the presence of two components, i.e. a core and a halo. By assuming a correspondence between the measured core spatial component and the lower divergence angular component of the beam, we obtain good agreement with previous values of the vertical emittance of the core beam estimated by JLAB. We have also placed bounds on the horizontal emittance of the core and both the horizontal and vertical emittances of the halo beam component. Further experiments are necessary to explain discrepancy between the previously estimated horizontal emittance and our measurement.

It is clear that a more precise methodology for determining the waist condition is needed to make more accurate measurements of the rms emittance. This may be attained by e.g. performing a quadrupole scan to achieve a minimum spot size before performing the OTRI divergence measurements.

The next phase of our experiments will employ an optical mask which will be used to spatially filter the light from a particular part of the beam's spatial distribution. This will enable us to individually separate and measure the size and divergence of the core and halo components.